\documentclass[useAMS,usenatbib,,usegraphicx]{mn2e}

\title[Unlocking the secrets of stellar haloes]{Unlocking the secrets of stellar haloes using combined star counts and surface photometry}
\author[Zackrisson et al.]{E. Zackrisson$^{1}$\thanks{E-mail: ez@astro.su.se}, R.S. de Jong$^{2}$, G. Micheva$^{1}$\\ 
$^{1}$Department of Astronomy, Stockholm University, Oscar Klein Center, AlbaNova, Stockholm SE-106 91, Sweden\\
$^{2}$Astrophysikalisches Institut Potsdam (AIP), An der Sternwarte 16, D-14482 Potsdam, Germany}

\begin{document}

\date{Accepted ... Received ...; in original form ...}

\pagerange{\pageref{firstpage}--\pageref{lastpage}} \pubyear{2011}

\maketitle

\label{firstpage}

\begin{abstract}
The stellar haloes of galaxies can currently be studied either through observations of resolved halo stars or through surface photometry. Curiously, the two methods appear to give conflicting results, as a number of surface photometry measurements have revealed integrated colours that are too red to be reconciled with the halo properties inferred from the study of resolved stars. Several explanations for this anomaly have been proposed -- including dust photoluminescence, extinction of extragalactic background light and a bottom-heavy stellar initial mass function. A decisive test is, however, still lacking. Here, we explain how observations of the halo of a nearby galaxy, involving a combination of both surface photometry and bright star counts, can be used to distinguish between the proposed explanations. We derive the observational requirements for this endeavour and find that star counts in filters $VI$ and surface photometry in filters $VIJ$ appears to be the optimal strategy. Since the required halo star counts are already available for many nearby galaxies, the most challenging part of this test is likely to be the optical surface photometry, which requires several nights of exposure time on a 4--8 m telescope, and the near-IR surface photometry, which is most readily carried out using the upcoming {\it James Webb Space Telescope}.  
\end{abstract}


\section{Introduction}
The stellar haloes of galaxies contain the fossil record of galaxy assembly, and observations of resolved halo stars in the Milky Way, Andromeda and other nearby galaxies \citep*[e.g.][]{Mouhcine et al.,Ibata et al.,Helmi et al.,Rejkuba et al.,Durrell et al., Tanaka et al.} have provided a wealth of information on such structures. However, stellar haloes can also be studied through surface photometry, i.e. observations of the integrated light from large numbers of unresolved stars within each system. 

Both techniques have their strong sides as well as shortcomings. Star counts can trace structures much further out into the halo, but this method is currently limited to the brightest halo stars (with the Milky Way halo as a notable exception) and to systems within $\sim 10$ Mpc. Surface photometry can be applied for more distant galaxies, is sensitive to less luminous halo stars and diffuse light from the interstellar medium, but works only for the relatively bright inner regions of halos. 

At the current time, the results from these two techniques are difficult to completely reconcile. A number of attempts to study the haloes of edge-on disk galaxies through optical and near-infrared (NIR) surface photometry have revealed halo colours that are too red to be consistent with the halo populations captured through the study of resolved stars \citep[e.g.][]{Sackett et al.,Lequeux et al. a,Lequeux et al. b, Rudy et al.,James & Casali,Lequeux et al. c,Zibetti et al.,Zibetti & Ferguson,Bergvall et al. b}. The most recent observations indicate that this red excess turns up at extremely faint surface brightness levels \citep[$\mu_i\approx 27$--29 mag arcsec$^{-2}$;][]{Zibetti et al.,Bergvall et al. b} and at wavelengths around the $i$-band ($\approx 7600$ \AA). Older measurements \citep{Rudy et al.,James & Casali} suggest that it may also continue into the near-infrared (12000--22000 \AA). On the other hand, surface photometry carried out at wavelengths shortward of the $i$-band seem to turn up nothing anomalous \citep[][]{Zibetti et al.,Jablonka et al.,Bergvall et al. b}.

It has been suggested that these `red halos' could be due to mundane observational effects like instrumental scattering \citep{de Jong} or flux residuals left over from an incomplete sky subtraction \citep{Zheng et al.,Jablonka et al.}. In some of the claimed detections, this may well be the case. However, \citet{Bergvall et al. b} correct for instrumental scattering (point spread function) effects and are still left with a pronounced red excess in their study of halos around stacked, edge-on low surface brightness disks. They also show that while the halos of these disks have extremely red colours, the disks themselves do not display any anomalous colours at the corresponding surface brightness levels. This strongly argues against a sky flux residual as the main culprit, since this would result in strange colours in both components. 

The red halo phenomenon is not necessarily confined to disk galaxies only. Near-infrared excesses have been reported in the faint outskirts of blue compact galaxies \citep{Bergvall & Östlin,Bergvall et al. a}, although \citet{Zackrisson et al. b} and \citet{Micheva et al.} have questioned the need for extreme stellar population properties to explain these. There are also older reports of extremely red optical colours in the outskirts of cluster elliptical galaxies \citep{Maccagni et al.,Molinari et al.}. However, no similar anomalies were seen in the recent study by \citet{Tal & van Dokkum} of the outskirts of 42000 stacked early-type galaxies (mostly located in galaxy groups). 

If one assumes that the red excess in the halos of edge-on disks is not due to instrumental scattering or sky flux residuals, this still leaves room for a number of possible explanations  -- including a bottom-heavy stellar initial mass function, diffuse light from the interstellar medium and extinction of extragalactic background light \citep{Zackrisson et al. a,Zackrisson et al. c,Bergvall et al. b}. These solutions can all give rise to similar halo colours. Given the extremely challenging task of improving current surface photometry measurements with existing telescopes, it has so far remained unclear if the degeneracy between these really can be broken. In this paper, we describe a method to discriminate between these scenarios using a combination of star counts and surface photometry. 

In Section.~\ref{explanations}, we review a number of explanations for the red excess that have previously been demonstrated to be broadly consistent with the available data. The principle of our proposed test is described in  Section~\ref{test}, and the observational requirements derived in Section~\ref{requirements}. Section~\ref{discussion} discusses a number of potential target galaxies that meet the distance requirements of the test. We also provide a worked example for the required exposure times in the case of the edge-on disk galaxy NGC\,4565, which is deemed to be one of the most suitable candidates. Our findings are summarized in Section~\ref{summary}.

\section{Possible explanations for the red excess}
\label{explanations}

\subsection{A bottom-heavy stellar initial mass function}
It has long been suspected that the red excess could be a signature of a stellar halo with an extremely bottom-heavy stellar initial mass function \citep{Lequeux et al. a,Rudy et al.,Zepf et al.,Zibetti et al.,Zackrisson et al. a,Bergvall et al. b}, i.e. a stellar halo with an abnormally high fraction of low-mass stars (with masses $<0.5\ M_\odot$). Due to the high mass-to-light ratios of such populations, solutions of this type were once thought to provide an explanation for the dark matter of galaxies. However, it is now well established that the dark matter is predominantly non-baryonic on cosmic scales \citep[e.g.][]{Komatsu et al.}. Surveys for microlensing events from Massive Astrophysical Compact Halo Objects (MACHOs) have also ruled out a dominant contribution to the dark halo of the Milky Way from objects in this mass range \citep[e.g.][]{Tisserand et al.}. On the other hand, up to $\approx 2/3$ of the cosmic baryons in the low-redshift Universe are still unaccounted for \citep[e.g.][]{Prochaska & Tumlinson}, and the missing-baryon problem may be even more severe on scales of individual galaxies \citep[e.g.][]{McGaugh et al.}. There is also a long-standing MACHO puzzle, in the sense that some halo microlensing surveys have turned up more events than can easily be attributed to known stellar populations \citep[e.g.][]{Alcock et al.,Calchi Novati et al.,Ingrosso et al.,Riffeser et al.}. Hence, it could be argued that there may be low-mass stars hiding in the halo that contribute to the missing-baryon problem and/or to the MACHO microlensing event rates, even though they have little bearing on the dark matter.

The stellar initial mass function (IMF) has long been assumed to be universal, but recent observational studies based on unresolved stellar populations have given some support to the notion that it may vary as a function of cosmic time \citep{van Dokkum}, environment \citep{Hoversten & Glazebrook,Meurer et al.,Lee et al.} and star formation rate \citep{Gunawardhana et al.}. The strength of certain absorption features in giant elliptical galaxies have been interpreted as evidence of a bottom-heavy IMF \citep{van Dokkum & Conroy a}, whereas the same absorption lines are consistent with a standard IMF in the case of globular clusters \citep{van Dokkum & Conroy b}. Attempts to constrain the IMF in the field population of the Large Magellanic Cloud using star counts \citep[e.g.][]{Gouliermis et al.} have also yielded results seemingly inconsistent with the standard IMF \citep[e.g.][]{Kroupa,Chabrier}. 

Surveys for low-mass stars (subdwarfs) in the Milky Way halo  have produced no evidence
for a bottom-heavy IMF \citep[e.g.][]{Gould et al.}, but current subdwarf searches are limited to distances out to $\approx 40$ kpc from the Sun. If there were a second halo population with a bottom-heavy IMF and a density profile different from that of the hitherto known halo, this second population could have evaded detection in the subdwarf searches. Such a halo population could on the other hand turn up in surface photometry measurements of external galaxies (and thus explain the red excess), due to projected contributions from stars further away from the halo centre. Scenarios of this type were explored by \citet{Zackrisson & Flynn}, who used the existing subdwarf data to rule out all secondary halo populations that would be able to contribute significantly to the missing baryons. Moreover, a halo like this would only remain viable if the surface brightness of the red halo reported by \citet{Zibetti et al.} for stacked edge-on disks had been significantly overestimated. This could for instance happen because of instrumental scattering, which Zibetti et al. did not properly correct for. The \citet{Zackrisson & Flynn} constraints, which are much stronger than those currently produced by microlensing searches, could also be sidestepped if the low-mass stars in the second halo were all locked up in star clusters. This, on the other hand, would cause tension with the \citet{Bergvall et al. b} stacked halo observations, which favour small pixel-to-pixel variations in halo colours  (i.e. a rather diffuse and not too highly clustered halo).

In summary, there are no observations that firmly rule out a halo component with a bottom-heavy IMF, but existing microlensing data, subdwarf star counts and surface photometry measurements do impose strong constraints on the properties of such a population. Current data place the mass that can be locked up in such a population at a level that is too low ($\ll 10^{10}\ M_\odot$) to have any significant bearing on the missing baryon problem ($\sim 10^{10}\ M_\odot$), and definitely not on the dark matter problem ($\sim 10^{12}\ M_\odot$) on the scale of Milky Way-sized galaxies \citep{Zackrisson & Flynn}. 

\subsection{Diffuse light from the interstellar medium}
\label{diffuse}
Another possibility is that the red excess does not originate from the stars in the halo, but is instead caused by some process in the interstellar medium. Diffuse, low-density halo gas could be kept ionized by far-ultraviolet photons leaking from the disk or by shocks and would contribute emission lines and a nebular continuum to the integrated halo spectrum. This could indeed give rise to strange halo colours, but current predictions for superpositions of old, low-metallicity stellar populations and ionized halo gas are still unable to match the observed colours of red halos at low redshifts \citep{Zackrisson et al. a,Bergvall et al. b}. Due to the high emission-line equivalent widths of a purely nebular spectrum, the optical colours of ionized halo gas evolve dramatically with redshift, and the colours reported by \citet{Zibetti & Ferguson} for a halo around an edge-on disk galaxy at redshift $z=0.322$ can in principle be explained by a mixture of ionized gas and a normal stellar halo \citep{Zackrisson et al. a}. However, the halo colours of this particular target may also be severely affected by instrumental scattering \citep{de Jong}.  

If the red excess originates in the interstellar medium, dust emission seems a more plausible radiation mechanism. Mid-infrared dust emission has been detected in the halos of several edge-on disk galaxies \citep[e.g.][]{Irwin & Madden,Burgdorf et al.,Kaneda et al.}, but to account for the red $r-i$ halo colours reported by \citet{Zibetti et al.} or \citet{Bergvall et al. b}, the dust emission would need to extend into the optical $i$-band. In principle, this could be achieved by the dust photoluminescence phenomenon known as extended red emission (ERE). While the dust component responsible for the ERE has yet to be identified, ERE has been reported in a wide range of environments including reflection nebulae, HII regions, the halo region of the starburst galaxy M82 and the high-latitude diffuse interstellar medium and cirrus in the Milky Way \citep[for a review, see][]{Witt & Vijh}. ERE manifests itself as a wide ($\sim$ 1000 \AA) emission feature in the wavelength range from 6100--9500 \AA, where the peak wavelength appears to vary from object to object. As argued by \citet{Bergvall et al. b}, ERE cannot be excluded as the mechanism responsible for the red $r-i$ halo colours, but anomalous colours involving filters at longer wavelengths \citep[e.g.][]{Rudy et al.} would then require a second explanation. 

\subsection{Extinction of extragalactic background light}
Anomalously red colours (as measured by colour indices like $V-I$, $r-i$ or $V-K$) do not necessarily imply the presence of a {\it flux excess} in the longer wavelength filter (e.g. the $i$ band in the $r-i$ colour index), since red colours can also be produced by a {\it flux deficit} in the shorter wavelength band (e.g. the $r$ band in $r-i$). This could most readily be caused by dust reddening of the halo stars, but as shown by \citet{Zackrisson et al. a} and \citet{Bergvall et al. b}, the observed red halo colours do not lie along standard dust reddening vectors. Moreover, the reddening needed to explain the observed $r-i$ halo colours would require $A(g)\approx 1$--3 magnitudes of $g$-band extinction, which seems completely implausible at projected distances of $\approx 5$--10 kpc into the halo. 

However, as demonstrated by \citet{Zackrisson et al. c} and \citet{Bergvall et al. b}, there is one mechanism that could account for the observed halo colours while requiring no more than $A(g)\approx 0.03$--0.2 mags of optical extinction, which would be consistent with the halo opacities inferred from the reddening of background light sources \citep{Zaritsky,Menard et al.,McGee & Balogh}. This mechanism is based on the extinction of extragalactic background light (EBL), which gives rise to a systematic sky subtraction problem for surface photometry measurements at faint isophotes. While most of the sky flux originates from regions between the telescope and the haloes targeted by surface photometry observations, a small fraction -- the EBL -- comes from behind. Provided that the existing direct measurements of the optical EBL are correct \citep{Bernstein}, most of this light appears to be diffuse (i.e. unresolved with all existing instruments). Unlike the other components of the night sky flux, the EBL can be subject to extinction by dust present in the target halos. This invalidates an assumption adopted in all current surface photometry measurements, namely that the sky flux and the flux from the target objects are unrelated. This extinction of the EBL may lead to a slight decrease in the overall sky flux across the target halos, which is neglected when the sky flux level is estimated well away from the target objects. The net effect is a systematic, wavelength-dependent oversubtraction of the sky, with spurious colour gradients in the outskirts of extended targets as the likely result \citep{Zackrisson et al. c}. 

If the red halo colours can be demonstrated to be due to EBL extinction, this could provide interesting constraints on the EBL itself. For halo opacities in the $A(g)\sim 0.01$--0.1 range (in line with current measurements), the surface brightness of the optical EBL would have to be at a level similar to that suggested by the \citet{Bernstein} EBL measurements. This is considerably higher than what currently resolved galaxies can account for \citep[e.g.][]{Madau & Pozzetti}. Hence, this would imply that the origin of the optical EBL remains unknown, with potentially far-reaching implications for cosmology. 

\section{A decisive test}
\label{test}
\subsection{Relating surface photometry to star counts}
\begin{figure*}
\includegraphics[width=84mm]{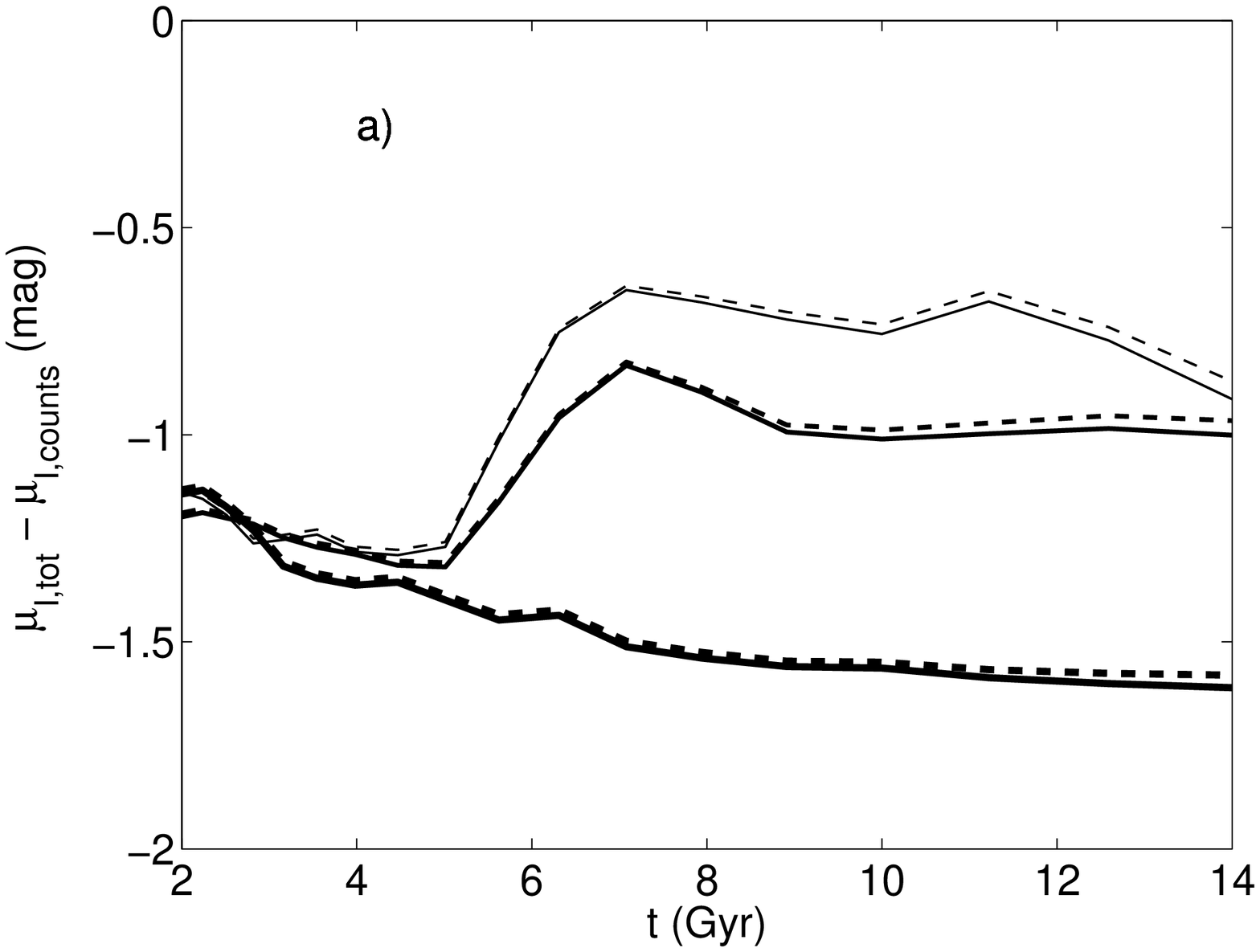}\includegraphics[width=84mm]{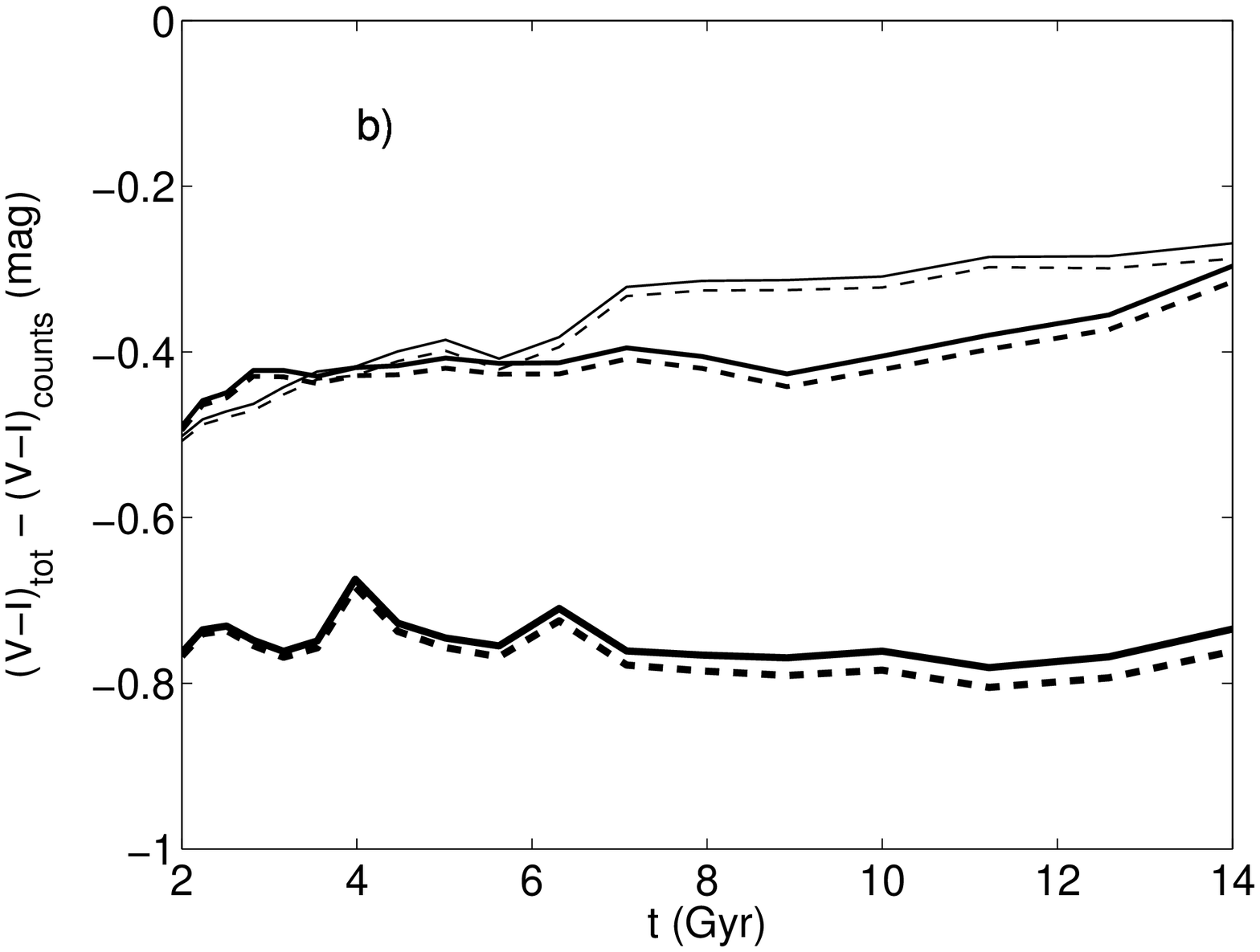}
\includegraphics[width=84mm]{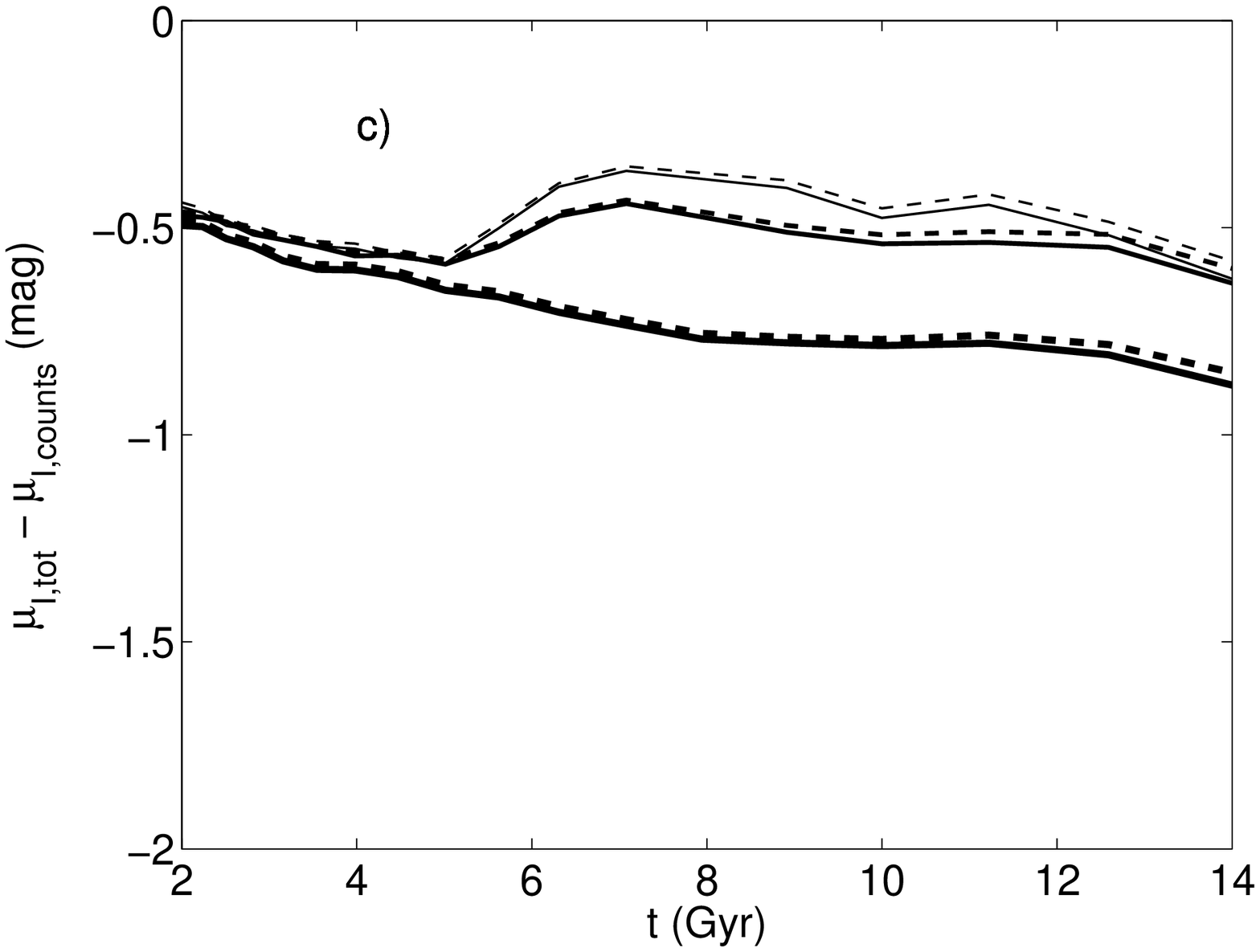}\includegraphics[width=84mm]{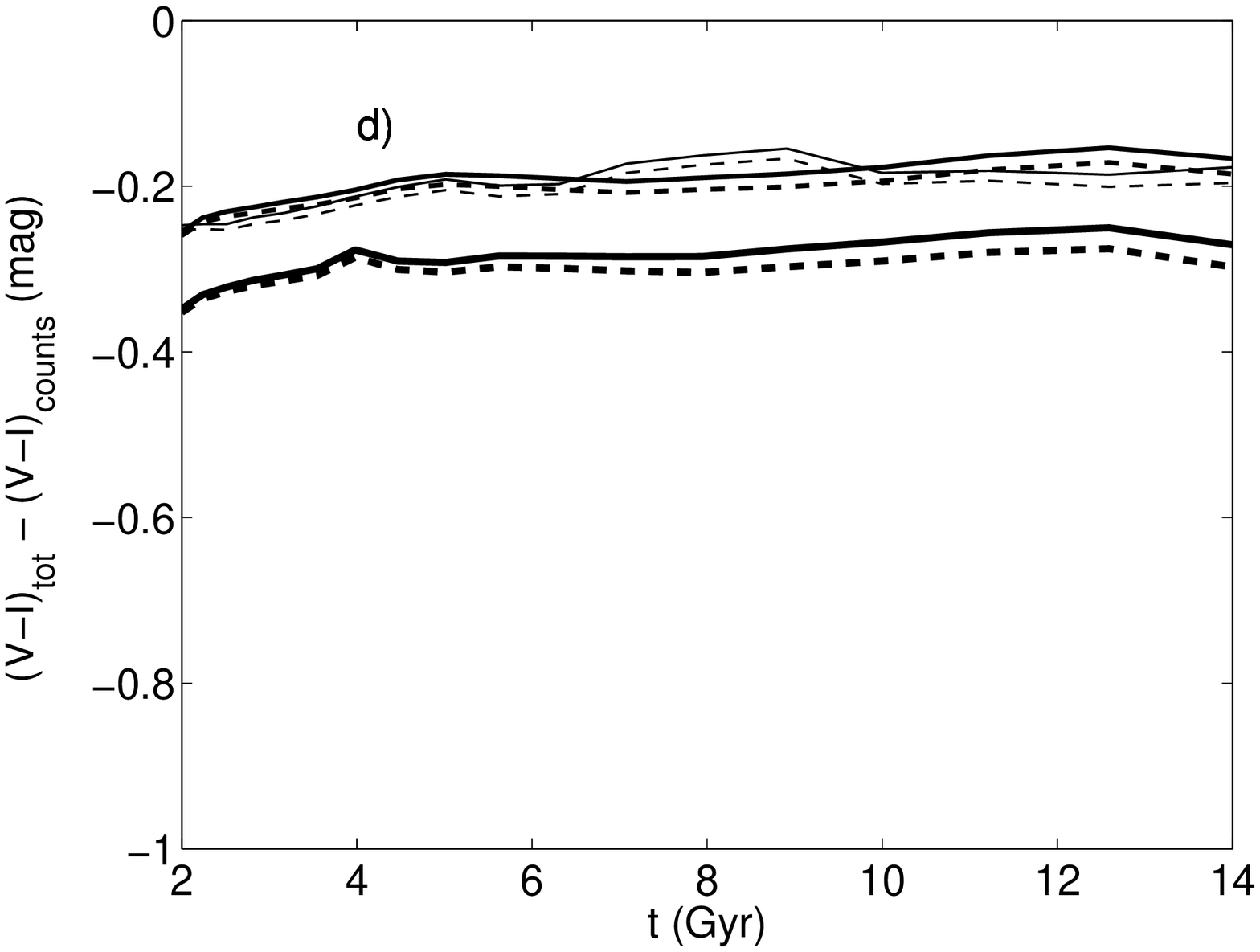}
\caption{Conversion between total surface brightness $\mu_\mathrm{total}$ (as measured through surface photometry) and the surface brightness due to resolved stars $\mu_\mathrm{counts}$ for an instantaneous-burst (i.e. single-age) stellar population. {\bf a)} The $I$-band conversion factor $\mu_\mathrm{total}-\mu_\mathrm{counts}$ as a function of age for metallicities $Z=0.0001$ (thin line), $Z=0.001$ (medium thick line) and $Z=0.008$ (thick line), in the case where stars are resolved 2.0 mag below the TRGB. Solid lines are for a \citet{Salpeter} IMF between 0.08 and 120 $M_\odot$, whereas the dashed lines corresponds to the \citet{Kroupa} IMF corrected for binary stars. The difference in $\mu_\mathrm{total}-\mu_\mathrm{counts}$ between these two IMFs is insignificant, but variations with age and metallicity variations are not. Still, the variations with age are small for old $\geq 6$ Gyr stellar populations, and the metallicity can typically be estimated from the colours of the TRGB, which means that uncertainties due to the metallicity can be corrected for. {\bf b)} The difference between the integrated $V-I$ colours of all resolved stars and the integrated $V-I$ colours of the entire population, $(V-I)_\mathrm{tot}-(V-I)_\mathrm{counts}$,  in the case where stars are resolved 2.0 mag below the TRGB. The lines have the same meaning as in panel a. The variations with age are relatively small ($\approx 0.2$ mag), but the variations with metallicity remain substantial ($\approx 0.5$ mag). {\bf c)} Same as b, but for counts reaching 4.0 mag below the TRGB. The uncertainty in $\mu_\mathrm{total}-\mu_\mathrm{counts}$ due to both age and metallicity variations is significantly reduced. {\bf d)} Same as b, but for counts reaching 4.0 mag below the TRGB. The sensitivity of $(V-I)_\mathrm{tot}-(V-I)_\mathrm{counts}$ to age variarions is very small, but metallicity variations can still produce significant colour shifts ($\approx 0.2$ mag).}
\label{conversion_ageZ}
\end{figure*}
\begin{figure*}
\includegraphics[scale=0.32]{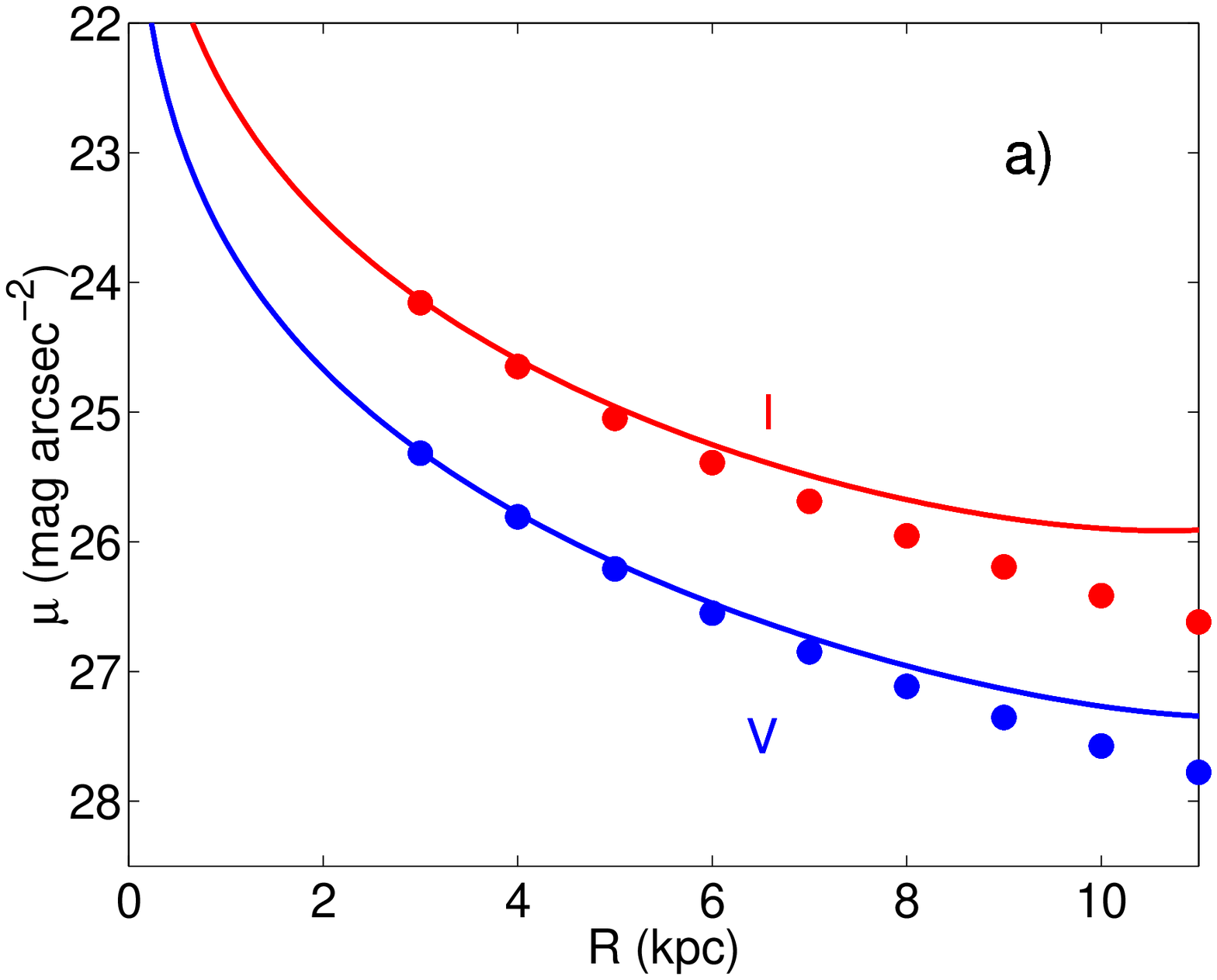}\includegraphics[scale=0.32]{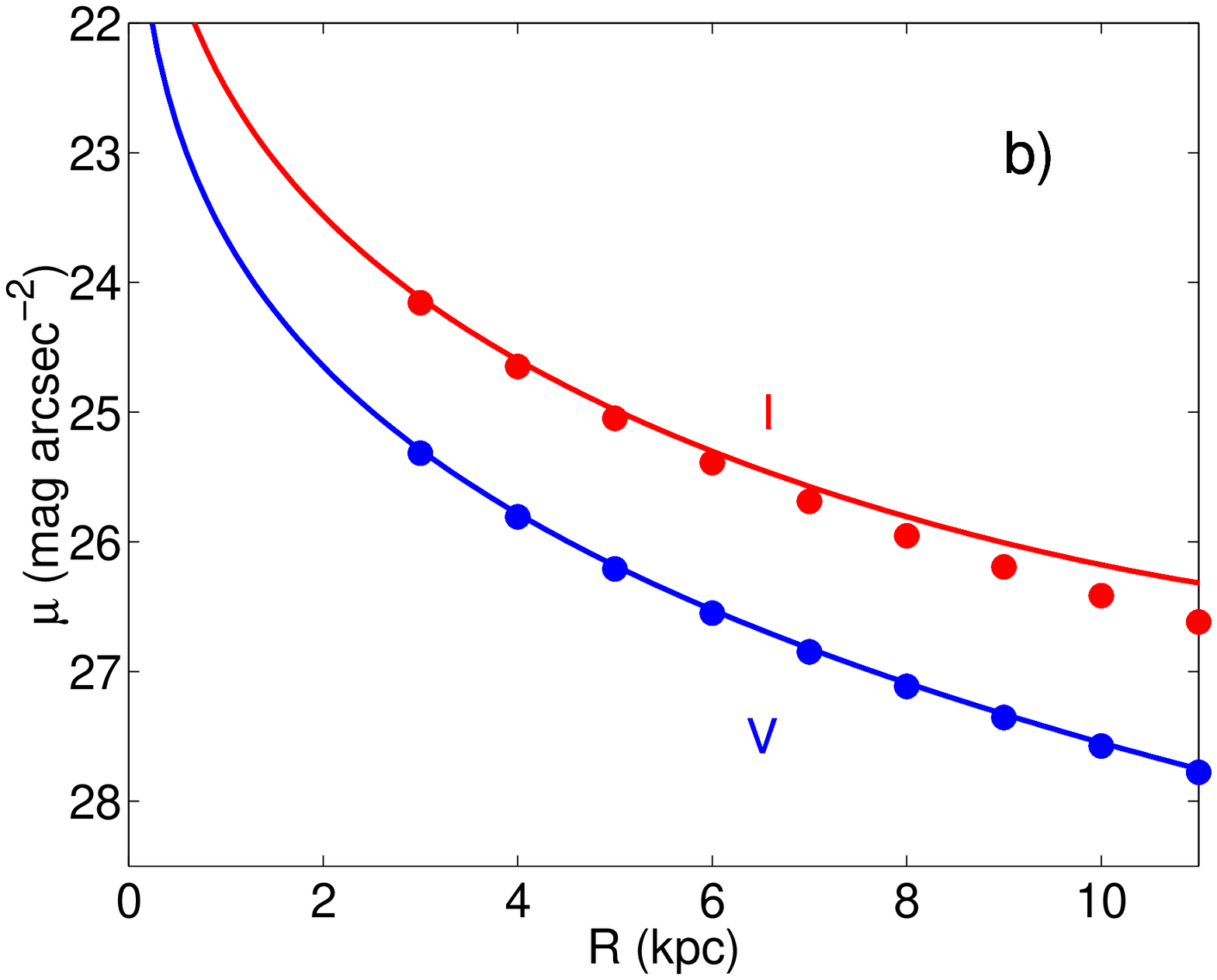}\\
\includegraphics[scale=0.32]{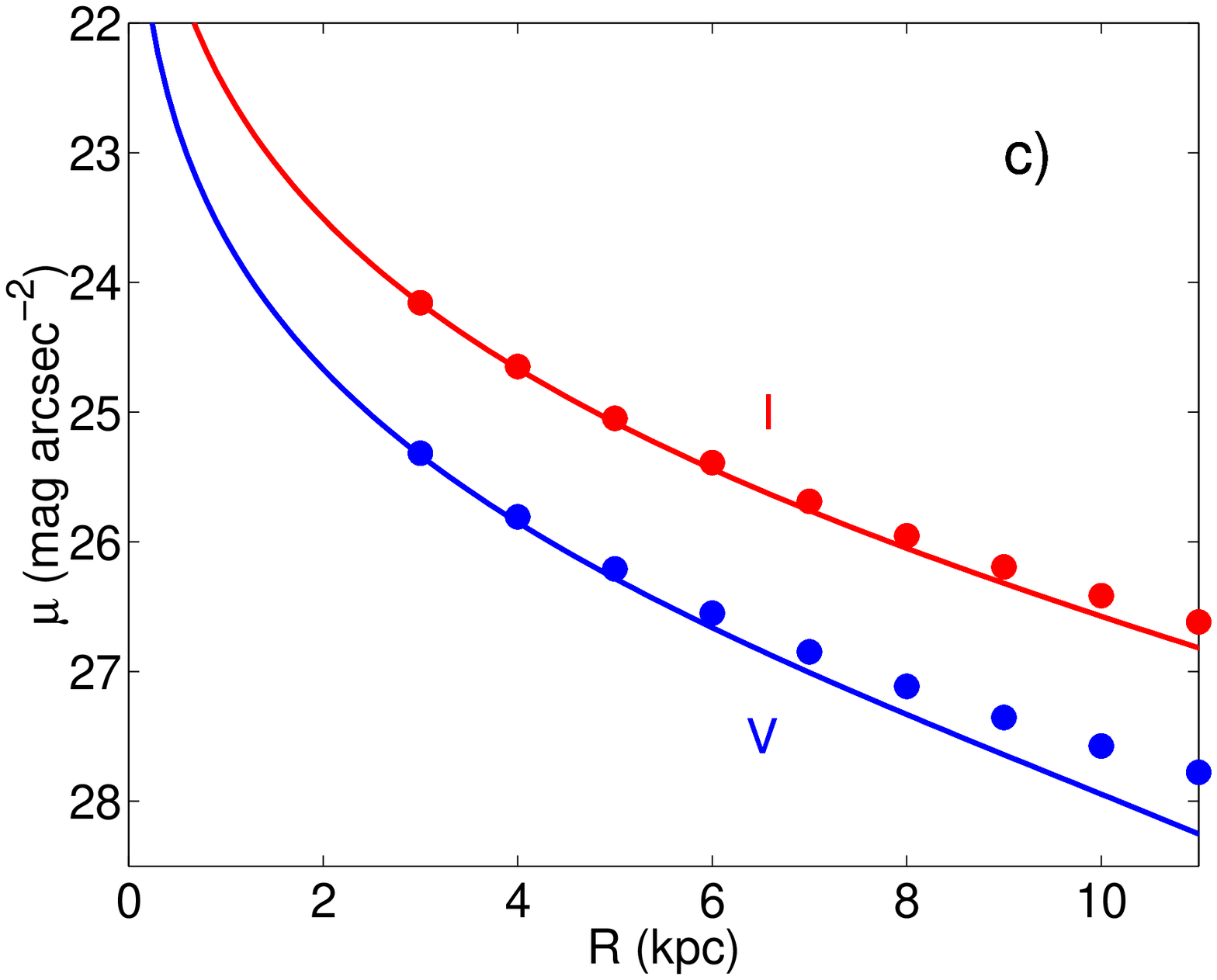}\includegraphics[scale=0.32]{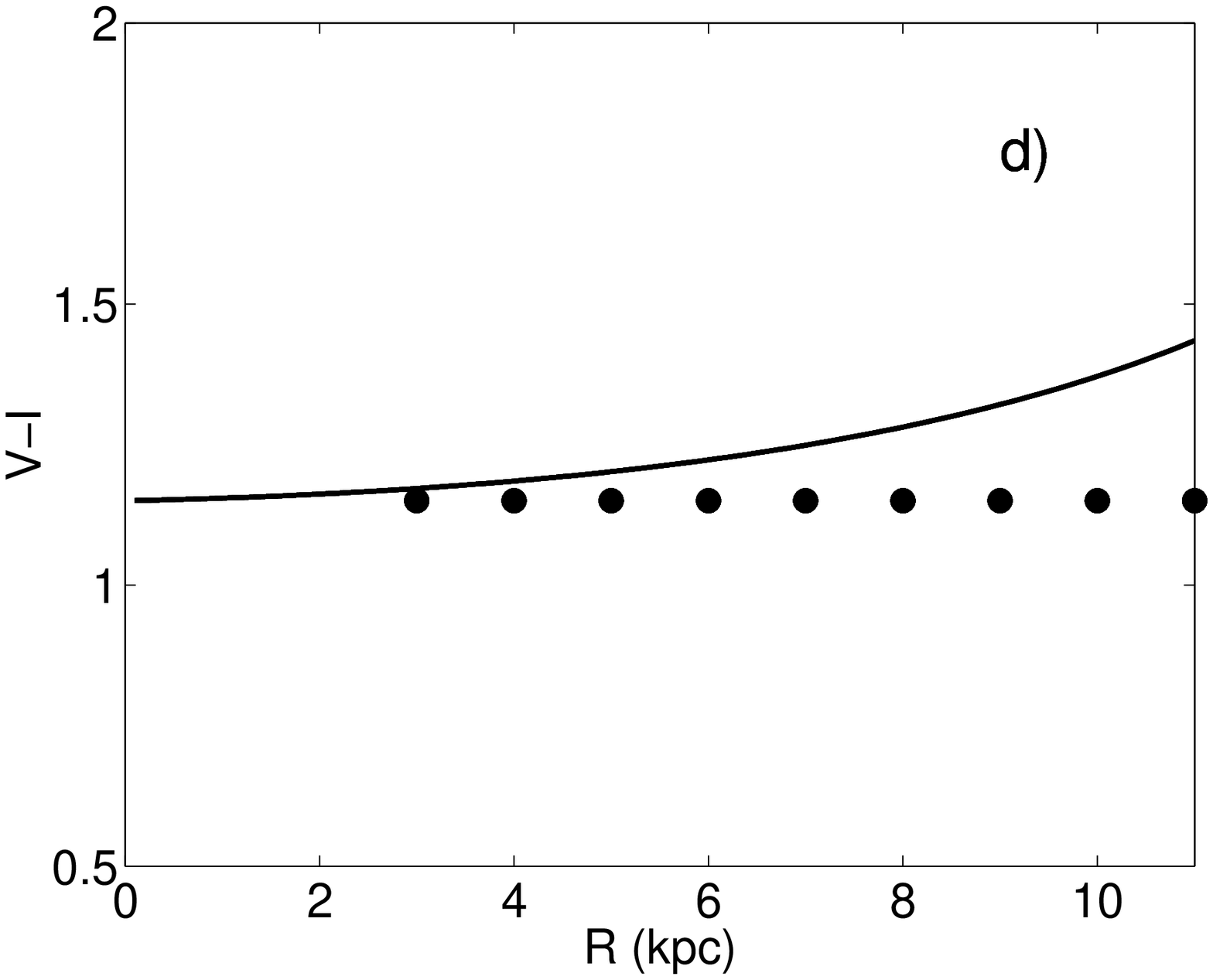}
\caption{Diagram illustrating how the possible explanations for red halo colours can be tested using a combination of star counts and surface photometry. {\bf a)} The halo surface brightness profile in $V$ (blue) and $I$ (red) as inferred from surface photometry (lines) and star counts (filled circles) in the case where the red halo colours are caused by excess flux in {\it both} the $V$ and $I$ bands, as would be expected from a stellar population with a bottom-heavy IMF. {\bf b)} Same in the case where the red halo colours is caused by dust photoluminescence (ERE), resulting in a flux excess limited to the $I$ band. {\bf c)} Same in the case where the red halo colours are caused by a larger flux deficit in the $V$-band than in $I$, due to EBL extinction. {\bf d)} The colour profile resulting from either of the situations depicted in a) and b) and c). See main text for additional details.}
\label{schematic}
\end{figure*}

In recent years, most attempts to trace the surface brightness profiles of stellar haloes have been based on direct star counts, i.e. by counting the number of bright halo stars (red giant branch stars, asymptotic giant branch stars and upper main sequence stars) within a certain area. 

The bright stars only account for a fraction of the overall halo light, which means that the star counts must be calibrated onto a surface brightness scale by making assumption about the stars below the detection threshold. Usually, this is achieved by simply shifting the star counts profile to match actual surface brightness measurements at bright isophotes \citep[e.g.][]{Irwin et al.,Barker et al.}. For an old stellar population with a reasonably normal IMF \citep[e.g.][]{Salpeter,Kroupa,Chabrier}, this procedure is fairly robust, as demonstrated in Fig.~\ref{conversion_ageZ}, where we present the conversion factor as a function of age, metallicity, IMF and depth of the star counts. These conversion factors are derived by integrating fluxes of stars along the \citet{Marigo et al.} stellar isochrones (with suitable IMF weights applied), and comparing the surface brightness $\mu_\mathrm{total}$ produced when the entire isochrone is used (the case relevant for surface photometry) to the surface brightness $\mu_\mathrm{counts}$ produced when only stars above a certain flux threshold are included (the case relevant for star counts). 

To illustrate how this conversion factor depends on the depth of the star counts, two cases are explored -- star counts reaching 2.0 mag below the tip of the red giant branch (TRGB) in the $I$-band (this corresponds to a detection threshold of $M_I\approx -2$) at an age of 10 Gyr in a $Z=0.001$ population (Fig.~\ref{conversion_ageZ}a \& b), and star counts reaching 4.0 mag below the TRGB (detection threshold $M_I\approx 0$) for the same age and metallicity (Fig.~\ref{conversion_ageZ}c \& d).  For simplicity, we assume that the star counts are sensitive to {\it all} stars brighter than the cutoff, and do not count stars in different evolved evolutionary stages separately. The TRGB provides a convenient reference point as its $I$-band magnitude of $M_I\approx -4$ is nearly independent of age and metallicity for any population older than 3 Gyr and more metal-poor than $0.5\times Z_\odot$, hence its use as a distance indicator of halo-like stellar populations.

While $\mu_\mathrm{total}-\mu_\mathrm{counts}$ shows substantial variations ($\approx 1$ mag) as function of both age and metallicity in the case of the shallower star counts (Fig.~\ref{conversion_ageZ}a), the variations with age are small ($\approx 0.2$ mag) in the age range relevant for old stellar haloes ($\geq 6$ Gyr) if one assumes that the metallicity is known. The $\mu_\mathrm{total}-\mu_\mathrm{counts}$ variations with age are moreover about the same in $V$ and $I$, so that the difference between the integrated colours of all resolved stars and the integrated colours of the entire population become no more than $(V-I)_\mathrm{tot}-(V-I)_\mathrm{counts}\approx 0.1$ at ages $\geq 6$ Gyr  (Fig.~\ref{conversion_ageZ}b). While this leaves room for substantial uncertainties in both $\mu_\mathrm{total}-\mu_\mathrm{counts}$ and $(V-I)_\mathrm{tot}-(V-I)_\mathrm{counts}$ due to radial metallicity variations, such metallicity gradients would result in radially varying colours for the TRGB, and can therefore be observationally tested for. For the deeper star counts (Fig.~\ref{conversion_ageZ}c \& d), both uncertainties are substantially smaller.

In summary, even if the stellar halos of galaxies contain substantial age gradients \citep[e.g.][]{Stinson et al.} or metallicity gradients \citep[e.g.][]{Carollo et al.}, the $\mu_\mathrm{total}-\mu_\mathrm{counts}$ uncertainties are small, provided that the integrated light comes from stars only, that the metallicity can be observationally assessed and that the IMF is of standard form. However, the reports of anomalously red halo colours based on surface photometry indicate that one (or several) of these tenets may be violated.

\subsection{Profile offsets}
The solutions to the red halo puzzle reviewed in Section~\ref{explanations} can be separated into two broad categories, depending on how the surface brightness profile inferred from surface photometry would differ from that inferred from star counts. Bottom-heavy IMFs and dust photoluminescence (ERE) would both decrease $\mu_\mathrm{total}-\mu_\mathrm{counts}$ (i.e. make the surface photometry profile brighter than the star counts profile), whereas EBL extinction would tend drive this quantity in the opposite direction. The principle is schematically illustrated in Fig.~\ref{schematic}, where we show how the halo surface brightness profiles based on star counts and surface photometry would deviate from each other in these three cases. Here, we have assumed that the star counts profiles have already been shifted to the level of the surface brightness profiles at bright isophotes (where a standard stellar population is assumed to dominate), so that the $\mu_\mathrm{total}-\mu_\mathrm{counts}$ offsets evident from Fig.~\ref{conversion_ageZ} have already been corrected for.

\begin{figure}
\includegraphics[width=84mm]{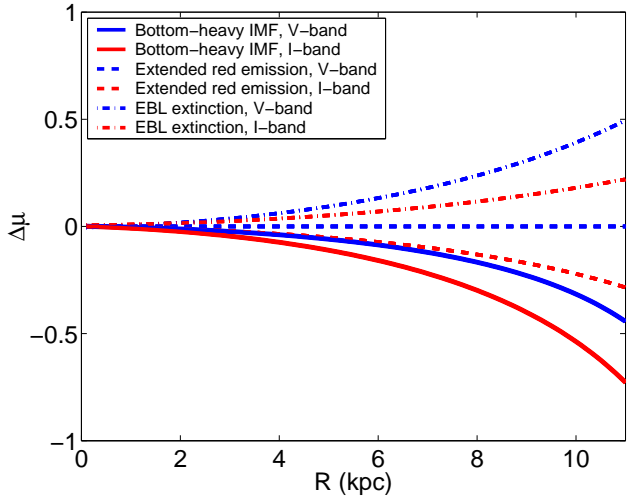}
\caption{The offset $\Delta\mu$ between the total surface brightness (as measured through surface photometry) and that inferred from star counts, in filters $V$ (blue lines) and $I$ (red lines) in the case where either a bottom-heavy stellar IMF (solid lines), ERE (dashed lines) or EBL extinction (dash-dotted lines) are responsible for pushing the integrated $V-I$ colours to anomalously red values (Fig.~\ref{schematic}d). A positive value on $\Delta\mu$ implies that the surface photometry profile lies faintward of the star counts profile, whereas a negative value implies that surface photometry profile lies on the brightward side. The three scenarios result in qualitatively different predictions in this diagram.}
\label{offsets}
\end{figure}

In the first case (Fig.~\ref{schematic}a), we assume that there is a halo component with a normal $V-I$ colour (similar to that of the stellar population hitherto detected in the Milky Way halo) that is slowly overtaken by a bottom-heavy IMF population with a much redder colour, thereby driving the integrated light measurements to extreme $V-I$ values at large projected radii. Since bright star counts fail to detect the increasing number of low-mass stars in the outer profile, the total surface brightness is shifted in the  brightward direction in both the $V$ and the $I$ bands, although more so in the $I$-band.

In the second case (Fig.~\ref{schematic}b), we assume that the same standard halo is affected by ERE in the $I$-band at large projected radii, thereby driving the integrated $V-I$ to the same extreme values as in Fig.~\ref{schematic}a. Whereas surface photometry may agree with the star counts in the $V$-band, the $I$-band surface photometry data would indicate a higher surface brightness than that inferred from star counts.

In the third case (Fig.~\ref{schematic}c), the colours of the normal halo component are pushed towards extreme values due to a failure to account for EBL extinction in the surface photometry measurements. Both the $V$ and $I$ surface photometry profiles would then indicate a surface brightness lower than that inferred by star counts. Here, the anomalously red integrated colours arise from a greater flux deficit in the $V$ band than in $I$. 

The model parameters have been chosen to produce the same integrated $V-I$ profile (Fig.~\ref{schematic}d) in all three scenarios, starting from $V-I=1.15$ (typical of an old, metal-poor stellar population) at small radii and reaching $V-I=1.45$ (anomalously red) at the outermost radius plotted. The offsets $\Delta\mu$ between the surface photometry and star counts profiles in Fig.~\ref{schematic} have been plotted in Fig.~\ref{offsets}. As seen, the directions and magnitudes of the $V$ and $I$ offsets are qualitatively different in these three cases. At the outermost data point, where a colour of $V-I\approx 1.45$ has been reached, the amplitude of the offset is 0--0.5 mag in the $V$-band and 0.2--0.75 mag in the $I$-band.

We argue, that by combining star counts and surface photometry for a single target in filters $V$ and $I$, one should at least be able to distinguish between an excess of red ($I$-band) flux (bottom-heavy IMF or ERE, as in Fig.~\ref{schematic}a and b) and a deficit of blue ($V$-band) flux (EBL extinction, as in Fig.~\ref{schematic}a) as the reason for the red halo colours. While Fig.~\ref{offsets} suggests that $V$ and $I$ data should also be able to distinguish between a bottom-heavy IMF from ERE, based on the lack of offset in the $V$-band in the latter case, this conclusion would not be as robust. While Fig.~\ref{offsets} certainly depicts a {\it possible} signature of ERE, it is not necessarily generic, simply because the colours of ERE are not sufficiently well-constrained. The ERE emission peak, which we have assumed to affect the $I$-band only, tends to coexist with a blue component of dust-scattered continuum radiation, whose relative contribution depends on the angle between the line of sight and the direction from which the ERE-fuelling ultraviolet photons are originating (in our case supposedly the disk). If this diffuse continuum is sufficiently strong, the offset $\Delta\mu$ between the surface photometry and star counts profiles would become non-zero in the $V$ band as well, thereby producing a situation similar to that of a bottom-heavy IMF (Fig.~\ref{schematic}a).

However, if the outcome favours an excess of red flux ($\Delta\mu_V$ and $\Delta\mu_I$ negative; $|\Delta\mu_I|>|\Delta\mu_V|$ in Fig.~\ref{offsets}) over a lack of blue flux ($\Delta\mu_V$ and $\Delta\mu_I$ positive; $|\Delta\mu_I|<|\Delta\mu_V|$ in Fig.~\ref{offsets}) as the reason for anomalous halo colours, a bottom-heavy IMF can be distinguished from ERE by additional surface photometry data in a filter at longer wavelengths ($J$, $H$ or $K$). In the case of a bottom-heavy IMF, the flux excess should grow stronger towards longer wavelengths, but not in the case of ERE, since this emission peak is limited to the wavelength range 6100--9500 \AA{}. Due to observational limitations (see Section~\ref{requirements}), the $J$ band appears to be the most suited for this endeavour. 

There is admittedly one possibility that is not covered by the scenarios presented in Figs.~\ref{schematic} and \ref{offsets}. Let's assume that the seemingly abnormal halo colours are caused by a failure of the models to adequately reproduce some bright, red phase of stellar evolution. This omission would need to be present in all of the many models already applied to the red halo problem \citep[e.g.][]{Bergvall et al. b}, and since the halo colours are considerably redder than the observed total colours of old systems like elliptical galaxies or globular clusters \citep{Zibetti et al.} it is perhaps more likely that there is a problem with the models at low to intermediate ages. If the integrated light of some region of the halo would be dominated by a relatively young, red component, the actual value of $\mu_\mathrm{total}-\mu_\mathrm{counts}$ could be closer to zero than predicted in Fig.~\ref{conversion_ageZ}, since a larger fraction of the integrated light would come from resolved stars. Provided that this happens at a radius greater than that at which the $\mu_\mathrm{total}$ and $\mu_\mathrm{counts}$ are normalized to the same value, this would cause the star counts (dots) to rise above the surface photometry profile (solid line) in Fig.~\ref{schematic}. Unlike the case for EBL extinction (Fig.~\ref{schematic}c),  the rise in the star counts would be more prominent in $I$ than in $V$ ($\Delta\mu_V$ and $\Delta\mu_I$ positive; $|\Delta\mu_I|>|\Delta\mu_V|$ in Fig.~\ref{offsets}), because of the very red colours of these bright objects. If this is accompanied by colour excess in the surface photometry measurements, this would be a tell-tale signature that the orgin of the abnormal halo colours should be sought among the resolved objects. However, a situation of this type is in conflict with the small difference between mean and median halo stacks measured by \citet{Bergvall et al. b}, which suggest that the sources responsible for the red halo colours are smoothly distributed throughout the halo and not concentrated to a few bright pixels. A sudden excess of very red, bright objects at some radius should also be directly detectable in the halo colour-magnitude diagram.

The surface brightness profiles depicted in Fig.~\ref{schematic} have been generated assuming that the surface brightness profile of the standard halo, as measured by star counts (corrected for the $\mu_\mathrm{total}-\mu_\mathrm{counts}$ offsets in Fig.~\ref{conversion_ageZ}) is described by a Sersic profile, with a constant $V-I\approx1.15$ at all radii. This colour is that of a 10 Gyr old, $Z=0.008$ population with a Salpeter IMF and an exponentially declining star formation rate $\tau=1$ Gyr ($\mathrm{SFR}\propto\exp(-t/\tau)$) predicted by integrated the \citet{Marigo et al.} isochrones and reweighting the timesteps according to the star formation history. Due to the relatively high metallicity assumed, this represents an upper limit on the $V-I$ colour that any normal halo population can be expected to display. The plotted star counts profile is described by a Sersic profile:
\begin{equation}
\mu_V=\mu_{V,0}+\frac{2.5b}{\log(10)}\left(\frac{r}{r_\mathrm{eff}}\right)^{1/n},
\end{equation}
with Sersic index $n=5$, $\mu_{V,0}=17$ mag arcsec$^{-2}$, $b=2n-0.324$ and $r_\mathrm{eff}=10.0$ kpc, resulting in an $I$-band profile similar to that derived by \citet{Zibetti et al.} in the polar direction away from their stacked edge-on disks. 

The effects of EBL extinction on surface photometry (blue and red solid lines in Fig.~\ref{schematic}c) have based on the diffuse EBL levels derived in \citet{Zackrisson et al. c}, a Milky Way extinction law and a radially constant $V$-band halo opacity of $A_V=0.05$ mag. The corresponding surface photometry prediction in the case of ERE (blue and red solid lines in Fig.~\ref{schematic}b) has been derived assuming that the diffuse ERE flux is limited to the $I$-band, while requiring the same colour gradient (shown in Fig.~\ref{schematic}d) as the EBL extinction scenario. The bottom-heavy IMF scenario prediction is based on the assumption that the standard halo at large radii is gradually overtaken by a second halo population with $V-I=1.5$ , as would be expected for a $Z=0.008$, 10 Gyr old $\tau=1$ Gyr population with an IMF slope $\alpha=4.50$ ($\mathrm{d}N/\mathrm{d}M\propto M^{-\alpha}$ throughout the mass range 0.08--120 $M_\odot$) -- one of the models discussed by \citet{Bergvall et al. b}. Based on the same \citet{Marigo et al.} isochrones as we use for the standard halo, we predict that the difference between the $\mu_\mathrm{total}-\mu_\mathrm{counts}$ offsets of the standard and bottom-heavy IMF halos are $\Delta\mu_V\approx-0.75$ and $\Delta\mu_I\approx-1.1$ in this case. The scaling of the relative fluxes of two components are then determined by the requirement of producing the same integrated V-I gradient (Fig.~\ref{schematic}d) as the EBL extinction and ERE scenarios.
 
While these choices of model parameters and profiles may seem arbitrary, we stress that our strategy for pinning down the mechanism responsible for the red excess is insensitive to the details of the adopted surface brightness or extinction profiles. Instead, the proposed test is based on the {\it offset} between $\mu_\mathrm{total}$ and $\mu_\mathrm{counts}$ over the limited isophotal range where a) surface photometry has revealed anomalous colours, and b) both $\mu_\mathrm{total}$ and $\mu_\mathrm{counts}$ can be measured (see Sect 4.1 and 4.2). In this case, it is only the typical values of $\mu_\mathrm{total}$, $\mu_\mathrm{counts}$ and extinction within this region of the halo that matter. The appearance of the profiles outside this region are not important. While the well-known structural components of galaxies (thin disk, thick disk, bulge and stellar halo) can be combined in endless ways to produce the same surface brightness level at a given radius, the offset between $\mu_\mathrm{total}$ and $\mu_\mathrm{counts}$ at that point still remains a sensitive probe of exotic ingredients entering the mix (Fig.~\ref{offsets}). 

\subsection{Choice of filters}
Even though filters other than $V$ and $I$ can in principle be considered for the optical part of these measurements, these filters appear to be the most suitable, both because similar passbands are routinely used in halo observations based on star counts and also because surface photometry can be pushed to very faint isophotes at these wavelengths. While the $R$ filter may seem more attractive than $I$ for doing surface photometry (due to a fainter sky background and less fringing), this strategy -- used in the halo measurements by \citet{Jablonka et al.} -- would introduce additional ambiguities since a red excess detected in $V-R$ could be caused by nebular emission (see Sect.~\ref{diffuse}) due to the location of the H$\alpha$ emission line in the $R$ filter. It is moreover unclear if the red excess detected by \citet{Zibetti et al.} and \citet{Bergvall et al. b} in $r-i$ really would be detectable in $V-R$ (even after correcting for the median redshifts of $z=0.05$--0.06 of these stacked samples). As far as we know, the only paper that lends any support for an anomalous $V-R$ colour in the halo of an edge-on disk is that of \citet{Rudy et al.} on NGC 5907. However, Rudy et al. arrive at their $V-R$ profiles by compiling inhomogeneous data from the literature. While their $V$-band profiles \citep[from][]{Lequeux et al. b}  differ on the two sides of the halo, the $R$-band profiles \citep[from][]{Sackett et al.} have been averaged over both sides of the halo. If data on one side of the galaxy is unreliable, this procedure may result in spurious effects. Indeed, \citet{Lequeux et al. c} note that the processing of the data on the west side of this galaxy (where a bright star happens to be located) is more troublesome than on the east. They also find that the $V-R$ colour they derive by combining their own $BVI$ measurements with the $R$-band profiles from \citet{Sackett et al.} produce $V-R$ colours for the halo that are seemingly inconsistent with $B-V$ and $V-I$ at the same radius.

\section{Observational requirements}
\label{requirements} 
The need to combine surface photometry data with star counts imposes a number of constraints on the observational strategy and the selection of suitable targets. 

Optical/near-infrared surface photometry measurements can easily trace the light distribution in the inner regions of galaxies, but become increasingly difficult as one moves outwards, towards fainter isophotes. The depth to which measurements of this type can be pushed is ultimately limited by one's ability to correct for instrumental scattering \citep{de Jong,Slater et al.} and to subtract off the sky with sufficient accuracy \citep{Zheng et al.,Mihos et al.,Jablonka et al.}. Star counts can be extended to much fainter isophotes, where the limit is set by contaminants due to foreground stars and background galaxies. However, star count profiles do not extend to the centre of the halo because of crowding problems at high surface brightness levels, as illustrated by the schematic surface brightness profiles included in Fig.~\ref{schematic}. To implement the proposed observational strategy, the overlap of the spatial regions for which both surface photometry and star counts can be carried out needs to be maximized. In the following, we only consider edge-on disk galaxies as potential targets, since the detection of a genuine stellar halo will be more ambiguous for other morphological types and orientations.

\subsection{Surface photometry}
\label{surfphot}
Even for the haloes of disk galaxies, there are substantial variations in the reported halo surface brightness at which anomalously red colours turn up. Using the \citet{Marigo et al.} model, we predict that a normal, old (age $\geq 6$ Gyr) metal-poor ($Z<0.008$), dust-free stellar population with a normal IMF should display $V-I\approx 0.9$--1.2. In the case of NGC\,5907, the most famous case of a red halo (see Sect.~\ref{discussion}), the observed colour profile reaches $V-I>1.2$ at $\mu_I\approx 26$ mag arcsec$^{-2}$ \citep{Lequeux et al. b,Lequeux et al. c}. 

In the haloes of stacked high surface brightness \citep{Zibetti et al.} and low surface brightness \citep{Bergvall et al. b} disks, the anomalously red $r-i$ colours are reported at $\mu_i\approx 26.7$ mag arcsec$^{-2}$ and $\mu_i\approx 28$--29 mag arcsec$^{-2}$ respectively. Converting the latter $i$-band results into the $I$-band requires knowledge of the spectral energy distribution of the halo in the $\approx 6500-11000$ \AA\ wavelength range, but the conversion term is -- at least for a halo dominated by direct star light -- not very sensitive to the stellar population properties adopted. For a metal-poor ($Z\leq 0.008$), old $\sim 10$ Gyr population with a standard IMF, the conversion is $i-I\approx 0.5$, whereas bottom-heavy IMFs of the type required to fit the $gri$ halo data \citep{Zackrisson et al. a,Bergvall et al. b} give $i-I\approx 0.6$. Adopting $i-I\approx 0.5$, these $i$-band isophotal levels convert into $\mu_I\approx 26.2$ mag arcsec$^{-2}$ and $\mu_I\approx 27.5$--28.5 mag arcsec$^{-2}$ respectively. 

Hence, while the \citet{Zibetti et al.} halo turns red at isophotes similar to those of NGC\,5907, the \citet{Bergvall et al. b} halo does not turn red until it is traced 1.5--2.5 mag fainter. This could be related to the different intrinsic properties (luminosity and disk surface brightness) of the sample galaxies, but at the current time, this remains unclear, since all of these data have not been subject to the same tests for systematical problems (instrumental scattering effects and sky subtraction residuals).

To cover all possibilities, the surface brightness measurements should ideally reach $\mu_I\approx 28.5$ mag arcsec$^{-2}$.  To robustly rule out or confirm a red excess, the $V$-band measurements should therefore ideally reach $\mu_V\approx 29.7$ mag arcsec$^{-2}$ (corresponding to $V-I=1.2$). To get a meaningful $V-I$ measurement (error $\sigma(V-I)\leq 0.15$), the $\sigma_V$ and $\sigma_I$ errors on the surface brightness measurements at these isophotes should not exceed 0.1 mag. Optical surface photometry on single targets have been pushed to similar levels before \citep[e.g.][]{Barton & Thompson,Zheng et al.,Jablonka et al.}, albeit not with an error this small. 

If an anomalously red $V-I$ colour is indeed detected in the halo, and the test outlined in Section~\ref{test} favours bottom-heavy IMF or ERE solutions over EBL extinction, it may become necessary to add near-infrared data as well. Since surface photometry measurements of halos are largely limited by the brightness of the night sky, and since the night sky becomes progressively brighter towards longer wavelengths, the observational prospects for matching the optical ($V$ and $I$) halo measurements with near-infrared data appear to be maximized in the $J$-band. A normal, old (age $\geq 6$ Gyr) metal-poor ($Z\leq 0.008$), dust-free stellar population with a normal IMF should display $V-J\approx 1.5$--2.0 \citep[][]{Marigo et al.}, whereas a bottom-heavy IMF would generate a $V-J$ redder than this (i.e. $V-J>2.0$). To be able to rule out or confirm an anomalously red  $V-J$ colour, the $\mu_V\approx 29.7$ measurements would then have to be complemented with $\mu_J\approx 27.7$ mag arcsec$^{-2}$ data (allowing a $J$-band detection if $V-J\geq 2.0$). On the other hand, if the red excess turns up already at $\mu_V\approx 27.2$ mag arcsec$^{-2}$ and $\mu_I\approx 26.0$ mag arcsec$^{-2}$ as in the case of NGC 5907, one could get away with $\mu_J\approx 25.2$ mag arcsec$^{-2}$ data.  Once again, the error should ideally not exceed $\sigma_J\approx 0.1$ mag at this isophote. 

\citet{Noeske et al.} have demonstrated that
groundbased $J$-band surface photometry can reach $J\approx 25$ mag
arcsec$^{-2}$ with a 3.5 m telescope. Assuming that shot noise dominates the
observations, one should be able to push the errors down to the required 0.1 mag by
scaling up to a bigger groundbased telescope, using longer exposure times and integrating over a larger halo region. We estimate, that to reach this level while maintaining $\sigma_J\approx 0.1$ mag with the {\it HAWK-I} instrument at the ESO
8.2m {\it Very Large Telescope}\footnote{www.eso.org}, one would require
about 3 hours of observing time (see Appendix A for details). 

However, it is unclear if observations at these isophotal levels really are
photon-limited, since the J band sky is dominated by OH emission and water
vapor absorption, both of which vary on both small and large spatial and
temporal scales. To reliably push the surface photometry faintward of
$\mu_J\sim 25$ mag arcsec${}^{-2}$, one would therefore need to resort to
space-based observations. Here, the upcoming {\it James Webb Space Telescope (JWST)} \footnote{www.jwst.nasa.gov}, tentatively scheduled for launch in 2018, may be one of the most suitable choices. When integrating over 0.37 arcmin$^2$, i.e. a 10\arcsec\ strip across the field of view ($2.2\arcmin \times 2.2\arcmin$) of the JWST NIRCam instrument, we find that a total of 2 hours in sky-chopping mode should be sufficient to reach the ideal depth of $\mu_J\sim 27.7$ mag arcsec${}^{-2}$. The details of these estimates can be found in Appendix A. In principle, data of this type can also be used to extend the star counts into the $J$-band, since the angular resolution of the {\it JWST} is expected to be superior to that of {\it HST} in this wavelength range.

\subsection{Star counts}
\label{starcounts}
Star counts in the haloes of nearby galaxies can readily be extended to surface brightness levels beyond $\mu_V\approx 30$ mag arcsec$^{-2}$. However, at bright levels, substantial crowding problems may occur. To implement the proposed test, one would need the star counts in $V$ and $I$ to trace the surface brightness profile from regions where no red excess is detected and into regions where it is. Since the red excess may turn up at isophotes as bright as $\mu_V\approx 27.2$ mag arcsec$^{-2}$ and $\mu_I\approx 26$ mag arcsec$^{-2}$ (see Sect.~\ref{surfphot}), one would ideally like to start the star counts in regions at least two magnitudes brighter than this, i.e. at levels around $\mu_V\approx 25$ mag arcsec$^{-2}$ and $\mu_I\approx 24$ mag arcsec$^{-2}$. 

\begin{figure}
\includegraphics[width=84mm]{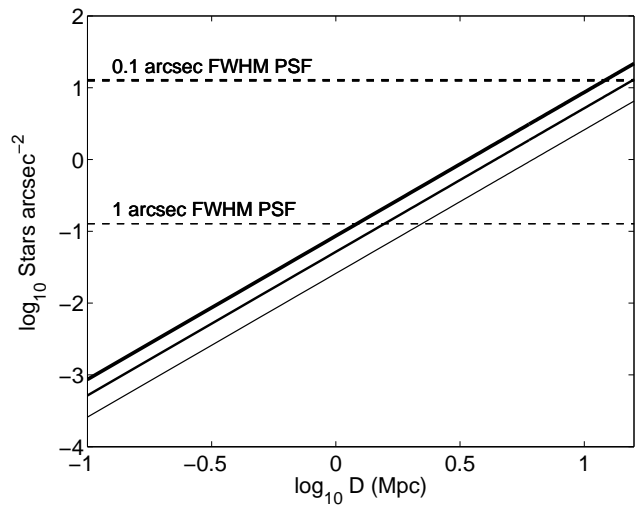}
\caption{The number of resolved stars per arcsec$^{2}$ as a function of distance for an old, metal-poor stellar population at $\mu_V\approx 25$ mag arcsec$^{-2}$. The diagonal lines correspond to the number of stars per arcsec$^{2}$ when the star counts are able to resolve stars 1 mag (thin solid line), 2 mag (medium solid) and 3 mag (thick solid) below the TRGB in the $I$-band. The dashed horizontal lines correspond to the limits for where crowding is expected to become serious for star counts carried out from the ground (thin dashed) and with the HST (thick dashed). The requirement of being able to perform star counts at $\mu_V\approx 25$ mag arcsec$^{-2}$ without serious crowding problems limits the distance to the target halos to less than 1.6 Mpc ($\log_{10} D (\mathrm{Mpc})\leq 0.2$) for groundbased observations and 16 Mpc ($\log_{10} D (\mathrm{Mpc})\leq 1.2$) for HST observations (assuming data reaching 2 mag below the TRGB).}
\label{crowding}
\end{figure}

In Fig.~\ref{crowding}, we present the number of resolved stars per square arcsecond at $\mu_V\approx 25$ mag arcsec$^{-2}$ as a function of distance to the target halo. These estimates are based on ratios of resolved stars per solar luminosity of $N_\mathrm{stars}/L_{V,\odot}\approx 3\times10^{-4}$ (thin solid line), $6\times 10^{-4}$ (medium solid) and $1\times 10^{-3}$ (thick solid) for detection thresholds 1, 2 and 3 magnitudes below the TRGB in the $I$-band, based on a population synthesis model constructed using the \citet{Marigo et al.} isochrones at $Z\leq 0.008$ and age $\geq 6$ Gyr. Also included are two crowding limits, based on the assumption that crowding becomes severe once the number of resolved stars per resolution element exceeds $\approx 0.1$. The two limits correspond to groundbased observations (thin dashed line; assuming circular resolution element with diameter 1\arcsec) and observations with the Hubble Space Telescope (thick dashed; assuming circular resolution element with diameter 0.1\arcsec). These assumptions result in crowding limits in rough correspondence with the isophotal levels where the incompleteness becomes severe in the star counts of \citet{Irwin et al.} and \citet{Seth et al.} for groundbased and spacebased observations, respectively. 

For groundbased observations, the crowding limit at $\mu_V\approx 25$ mag arcsec$^{-2}$ is reached at distances of $D> 1.6$ Mpc (i.e. slightly beyond the Local Group), with only a weak dependence on the depth of the observations. For Hubble Space Telescope (HST) observations, the limit sets in at $D> 16$ Mpc. Since there are very few high-inclination disk galaxies within 1.6 Mpc from the Milky Way, the HST appears to be the best choice for the star counts at the current time. 

\section{Discussion}
\label{discussion}
The need to obtain matching star counts and surface photometry data requires a nearby target (Sect.~\ref{starcounts}), with a large angular halo diameter as a result ($\approx 5$--10$\arcmin$). Since the the HST currently represents the best option for the star counts (with a field of view of $3.3\arcmin \times 3.3\arcmin$ for the ACS instrument), the proposed test will realistically be limited to a small region of the halo. While this means that halo substructure (in the form of satellite galaxies and tidal streams) may well interfere with the proposed measurements, this should not affect the outcome of the proposed test, since such structures are expected to affect the star counts and surface photometry in similar ways. In fact, the modern view of a stellar halo is that of large collection of substructures that slowly dissolve into the diffuse component, which is created by both accretion events and in-situ star formation \citep[e.g.][]{Bullock & Johnston,Zolotov et al.}. At this time, it is unclear whether the red excess originates in the diffuse halo component, in halo substructure or in both.

Most of the massive ($V_{\rm rot}$$>$100 km/s), edge-on (inclination $>$ 85 degrees) disk galaxies that have reliable distance estimates that meet the distance criteria for $HST$ star counts ($D\leq 16$ Mpc) have already been targeted by the GHOSTS\footnote{Galaxy Halos, Outer disks, Substructure, Truncations, Star Clusters} survey \citep[][]{de Jong et al.,Radburn-Smith et al.}, which means that the star counts data are already available. To implement the proposed test, one therefore simply has to obtain matching surface photometry data. Below, we discuss a number of potential targets for which star counts are already available from GHOSTS or elsewhere --  NGC 5907, NGC 7814, NGC 3109 and NGC 4565. We argue that NGC 4565 is likely to be the best target for this endeavour, and present a worked example of the required exposure times on telescopes of various sizes. As will be demonstrated, a 4--8 m class telescope represents the most practical option for the $V$ and $I$-band surface photometry.

The large diameters of these halos at the relevant isophotes ($\approx 5$--10$\arcmin$) imply that the surface photometry measurements either have to be carried out with wide-field imagers (e.g. Subaru/Suprime-Cam\footnote{http://www.naoj.org}, VISTA\footnote{http://www.vista.ac.uk/}, CFHT/MegaCam\footnote{http://www.cfht.hawaii.edu/Instruments/Imaging/MegaPrime/}) - or in sky-chopping mode. Both approaches are admittedly very challenging at the extremely faint isophotes where the red excess is likely to turn up and a combination of both may provide the best route in the long run. 

While wide-field imagers save observing time, one tends to have better
overview and control of the bias level and flatfielding accuracy in
single chip imaging with single amplifier readout. Additionally, the
single chip/amplifier setup avoids or at least significantly
reduces the effects of filter-dependent large-scale variations of the
pixel response, crosstalk and amplifier gain differences. On the other
hand,  due to the necessary proximity of the target galaxy, using a
single chip would imply that the level of the background sky cannot be
estimated from the target observations. Therefore chopping mode
observations are necessary in order to obtain separate imaging of the
sky. Chopping mode is not a traditional observing strategy in the
optical. However, it is routinely used to reach surface brightness levels $\sim 10000$ times below the sky in the near-IR regime \citep[e.g.][]{Cairos et al.,Noeske et al.,Vaduvescu et al.}, where the temporal sky variations are rapid and can easily result in
residual low frequency variations of the background in the
sky-subtracted frame. The optical sky is in contrast much more stable
and there is no obvious reason why chopping mode in the optical would
not be at least as successful as it is in the comparatively more
troublesome NIR regime. For this reason, our exposure time estimates are based on the assumption that the surface photometry observations are carried out in sky-chopping mode. 

\subsection{NGC\,5907}
The edge-on disk galaxy NGC 5907, at a TRGB distance of 16.8$\pm$0.8 Mpc \citep{Radburn-Smith et al.}, by far represents the most famous case of a claimed `red halo' \citep[e.g.][]{Sackett et al.,Lequeux et al. b,Rudy et al.,Lequeux et al. c,James & Casali,Zepf et al.}. As discussed in Sect.~\ref{surfphot}, the observed $V-I$ profile reaches an anomalous value already at $\mu_I\geq 26$ mag arcsec$^{-2}$ \citep{Lequeux et al. b,Lequeux et al. c}. Interest in this object declined with the discovery of what appeared to be the remnants of a disrupted dwarf galaxy on one side of the halo \citep{Shang et al.} and suggestions that this ring-like feature, in combination with contamination by foreground stars, could have resulted in the spurious halo detection \citep{Zheng et al.}. To some extent, this is a matter of semantics, since tidal features are nowadays regarded as ubiquitous features of stellar halos \citep[e.g.][]{Bullock & Johnston}. The details of how this tidal feature supposedly explains the extreme halo colours reported by others remain unclear, especially since the ring itself appears to have normal colours \citep{Zheng et al.}. 

While the colour anomaly of NGC\,5907 is well reported, its distance makes it a very difficult target for our test. To obtain accurate point source photometry 2 mags below TRGB would require at least 15 HST orbits per pointing. Several HST pointings would probably be required, as \citet{Martinez-Delgado et al. a} have reported a complex system of streams around this galaxy, and halo measurements on and off the streams would be required to determine the origin of the colour anomaly. The 6 orbit HST observation of the GHOSTS survey was barely deep enough to provide a TRGB distance to this galaxy and contains too few stars for a reliable luminosity profile. It is furthermore located poorly on the major axis near the disk truncation at the end of the disk and gives no clean halo measurement.

\subsection{NGC\,7814}
The edge-on galaxy NGC 7814, at a distance of 14.4$\pm$0.7 Mpc \citep{Radburn-Smith et al.} has also been argued to exhibit an anomalous $V-I$ colour in its outskirts \citep{Lequeux et al. a}. This object is extraordinary in the sense that the excess seemingly turns up at relatively bright isophotes ($\mu_I\approx 23$--24 mag arcsec$^{-2}$). No follow-up observations have apparently been undertaken. However, due to the very prominent bulge in this galaxy, and the fact that the colours were not measured in the polar direction but in the outer disk region, it is not clear which component of the galaxy is responsible for this. 

The prominent bulge and the relatively large distance make NGC\,7814 a difficult target for our test. The HST star counts suffer from crowding at about $\mu_I\approx$ 24 mag arcsec$^{-2}$ at this distance, exactly where the colour anomaly becomes significant if the \citet{Lequeux et al. a} measurements are correct. Extensive HST imaging data does exist for this galaxy with 4 ACS and WFC3 fields along the minor axis out to about 14 arcmin radius ($\sim$60kpc) and several more fields along the major axis. While absolute matching of the star counts and integrated light profile in the central region of the galaxy may not be possible, the relative profiles could definitely be compared over at least 3 magnitudes in surface brightness where \citet{Lequeux et al. a} claim a strong colour gradient.  
However, a cursory look at the pipeline-reduced Sloan Digital Sky Survey\footnote{www.sdss.org} mosaics (data release 8) does not give any obvious support for an anomalous or even rising $g-i$ colour at the distance where \citet{Lequeux et al. a} note a $V-I$ excess. 

\subsection{NGC\,3109}
NGC 3109, at a distance of $1.32\pm0.06$ Mpc \citep*{Hidalgo et al.}, represents one of the few high-inclination disk galaxies located sufficiently close to make groundbased star counts a realistic alternative for our proposed test. Star counts reaching $\approx 2$ mag below the TRGB in the $V$ and $I$ bands are already available in regions located in the direction perpendicular to the disk \citep{Hidalgo et al.}, allowing the surface densities of resolved stars to be traced $\approx 10\arcmin$ ($\approx 4$ kpc) into the halo (out to $\mu_V\approx 30$ mag arcsec$^{-2}$). There is, however, no reported colour anomaly in the halo of this galaxy, and its relatively low Galactic latitude ($b=23.1\deg$) makes surface photometry measurements prone to the effects of Galactic cirrus and scattered light from foreground stars.

\subsection{NGC\,4565}
NGC\,4565 is a near-perfect edge-on galaxy that is ideally suited for our research. It is at a distance of $D=11.9\pm 0.3$ Mpc as determined from the TRGB from {\it HST} imaging \citep{Radburn-Smith et al.}, which is near enough that the RGB population can be comfortably be measured with HST, while it has an angular diameter small enough ($D_{25}\sim$16 arcmin)  that a large fraction of its halo can be observed within a single image of a groundbased camera, allowing good flatfielding checks. NGC\,4565 has a high galactic latitude ($b=86.4\deg$) and therefore Galactic cirrus and scattered light from foreground stars will have minimal effect on the integrated light observations. There are no obvious tidal streams in existing deep images, but an H{\sc i} tidal bridge exists to a small dwarf galaxy to the North. However, no anomalous halo colours have been reported for this galaxy as of yet.

Existing {\it HST} images obtained with the {\it Advanced Camera for Surveys} ({\it ACS}) and {\it Wide Field Camera 3} ({\it WFC3} cameras as part of the GHOSTS survey with long exposure times ($>$7000 s)  in F606W (equivalent to $V$-band) and F814W ($I$).  So far 5 fields have been obtained, two of which along the minor axis, and one more observation with the {\it ACS} and {\it Wide-Field Camera 3} ({\it WFC3}) cameras working in parallel has been scheduled. The observations are deep enough to reach $\approx 2.0$ mag below the TRGB, which is sufficient for the proposed test. Starting a few kpc above the disk the color-magnitude diagrams are completely dominated by RGB and AGB stars with no young stars indicated. Even at 30 kpc above the midplane of the galaxy, the estimated contamination of unresolved background galaxies is still less than 2\%.

To match these {\it HST} star counts with $V$ and $I$-band surface photometry data, new ground-based observations would be required. Based on the \citet{Wu et al.} $i_{6660}$-band surface brightness profile (similar to the $R$ band), we estimate that the surface brightness level of the NGC\,4565 halo in the most distant {\it GHOSTS} field (at $\approx 6$\arcmin{ } from the disk) to be $\mu_R\approx28$ mag arcsec${}^{-2}$. This can be converted into corresponding $\mu_V$ or $\mu_I$ values, by assuming a shape for the observed halo spectrum in this wavelength range. The \citet{Marigo et al.} model predicts that an old and metal-poor stellar population with a standard IMF would display $V-R\approx 0.5$ \citep[see also the discussion in][]{Jablonka et al.} and $R-I\approx 0.5$, but if this were the case, there would be no red excess and the test would not be meaningful. \citet{Zackrisson et al. a} and \citet{Bergvall et al. b} present models for old stellar populations with bottom-heavy IMFs that are able to reproduce the anomalous $gri$ halo colours of stacked edge-on disks. These models predict $V-R\approx 0.8$, $R-I\approx 0.9$. EBL extinction models capable of fitting the $gri$ data can produce normal $g-r$ colours yet $r-i$ as red as those predicted by a bottom-heavy IMF \citep{Zackrisson et al. c}, suggesting $V-R\approx 0.5$, $R-I\approx 0.9$. However, since the $R$-band surface brightness is fixed, the observations become more challenging if $V-R$ is red (implying a fainter $V$-band profile) rather than blue (brighter $V$-band profile). In order not to produce estimates that are overly optimistic, we therefore adopt the more extreme estimates on both colours: $V-R\approx 0.8$ and $R-I\approx 0.9$. The $\mu_R\approx28$ mag arcsec${}^{-2}$ isophote may then correspond to $\mu_V\approx28.8$ mag arcsec${}^{-2}$ and $\mu_I\approx27.1$ mag arcsec${}^{-2}$. This is somewhat brighter than the ideal requirements derived in Sect.~\ref{surfphot}, but sufficient to uncover a red halo if it is similar to the stacked halo detected by \citet{Zibetti et al.}, $\mu_I\approx 26.2$ mag arcsec$^{-2}$. 

By integrating over 10\arcsec\ strips across this {\it GHOSTS} field ($3.3\arcmin \times 3.3\arcmin$), we estimate that one can reach the required levels ($\mu_V\approx28.8$ mag arcsec${}^{-2}$,  $\mu_I\approx27.1$ mag arcsec${}^{-2}$) with an error of $\sigma=0.1$ mag in approximately 2 hours for the $V$ band and 4 hours for the $I$-band with the 8.2m {\it VLT} and the {\it FORS2} instrument (see Appendix A for details). For comparison, the corresponding times on smaller telescopes (with diameters $D=4$, 2, 1 and 0.5m, respectively) would be $T=9$, 37, 148 and 592 hours for the $V$ band and $T=17$, 67, 269 and 1076 hours for the $I$ band. This programme would seem best suited for telescopes with diameters in the 4--8 m range, since smaller telescopes require excessive amounts of observing time. There is, however, otherwise nothing that precludes the use of smaller telescopes for measurements of this type. As demonstrated by \citet{Martinez-Delgado et al. a,Martinez-Delgado et al. b}, tidal features in halos have in fact been detected at similar isophotes using long observing campaigns at very small telescopes ($D\leq 0.5$m). 

Should the test tend to favour an excess of $I$ band flux over a deficit of $V$ band flux as the likely explanation for the anomalous halo colour, follow-up $J$-band observations may be required to distinguish between ERE and a bottom-heavy IMF. As argued in Sect.~\ref{surfphot}, a normal stellar halo should display $V-J\approx 1.5-2.0$. This could also be the $V-J$ colour expected in the case of ERE, since neither the $V$ or $J$ filters are expected to be strongly affected by this mechanism. To rule out a normal $V-J$ colour in the NGC\,4565 field, one may therefore need to reach as deep as the the $\mu_J=\mu_V-2.0$ isophote, which in this case corresponds to $\mu_J\approx 26.8$ mag arcsec$^{-2}$. The {\it JWST} appears to be the best option for this. The estimates provided in Sect.~\ref{surfphot} and Appendix A indicate that one should be able to reach this level in less than two hours. Despite having a relatively small field of view ($2.2\arcmin \times 2.2\arcmin$), the improved sensitivity of the {\it JWST/NIRCam} instrument compared to existing {\it HST} cameras will also allow resolved halo stars to be mapped across the halo, in just a fraction of the {\it HST }observing time spent on existing {\it GHOSTS} fields.

\section{Summary}
\label{summary}
Attempts to measure the integrated optical/near-IR colours of the stellar halos of galaxies through surface photometry have revealed halo colours that are at odds with the halo properties inferred from the study of bright, resolved halo stars. Several possible explanations -- a bottom-heavy stellar IMF, diffuse light from the interstellar medium and extinction of extragalactic background light -- have been proposed, but a test capable of differentiating between these is still lacking. We argue that these different scenarios can be tested by combining star counts in the $V$ and $I$ bands with similar surface photometry data, since the proposed explanations should result in different offsets between these profiles. When using the {\it HST} for the star counts, this test can be applied for galaxies up to $D\approx 16$ Mpc. Matching $V$ and $I$ surface photometry data can be obtained with existing ground-based telescopes. We discuss a number of potential edge-on disk galaxies for which halo star counts are already available, and argue that the NGC 4565 is likely to be the best target for this endeavour. 

\section*{Acknowledgments}
E.Z. acknowledges research grants from the Swedish National Space Board, the Swedish Research Council and the Royal Physiographical Society of Lund. This research has made use of the NASA/IPAC Extragalactic Database (NED) which is operated by the Jet Propulsion Laboratory, California Institute of Technology, under contract with the National Aeronautics and Space Administration. The authors would also like to thank the referee, Ken Freeman, for insightful comments which helped improve the quality of the paper.

\appendix
\label{appendix}
\section{Estimates of required exposure times}
The need to target galaxies at distances $\leq 16$ Mpc implies that the halos of large galaxies will measure many arcminutes, possibly even degrees, across (e.g. a halo with radius 50 kpc at a distance of 16 Mpc will have an angular diameter of 21.5\arcmin). While having a telescope field of view sufficiently large to cover the entire halo would certainly be an asset, the proposed test in principle requires only surface photometry observations of the part of the halo in which the star counts have been carried out. In the following, we assume that the star counts will be space-based, and adopt an area of $3.3\arcmin \times 3.3\arcmin$ (corresponding to the {\it HST/ACS} observations). Since wide-field imagers tend to be rare on the largest telescopes, we therefore base our exposure time estimates on the premise that the surface photometry needs to cover this area only.

The observational and reduction strategies will therefore involve chopping mode and consequent pair-subtraction of images. Since this is not standard practice for optical observations, optical exposure time calculators (ETCs) do not (unlike NIR ETCs) account for the increased noise that results from the pair subtraction. This implies that any exposure time estimated for on-source observations in the optical must be quadrupled to obtain the total required telescope time. For the duration of this paper all presented optical exposure times are total required times which include on-source and off-source observations and a further factor of 2 to compensate for the pair-subtraction. For HAWK-I, a NIR instrument, the total increase factor of the ETC estimated time is only 2.

\subsection{Optical surface photometry}
In the optical, ground-based telescopes should suffice since the atmosphere is almost transparent in this wavelength range. The estimated surface brightness level of the NGC\,4565 halo at the distance of the GHOSTS field at $\approx 6\arcmin$ or $\approx 20$ kpc away from the centre is $\mu_V\approx28.8$ mag arcsec${}^{-2}$ and $\mu_I\approx27.1$ mag arcsec${}^{-2}$. For each radial data point along the surface brightness profile, the integration area is assumed to have length $3.3$ arcmin parallel to the plane of the disk and width of $10$ arcsec in the perpendicular direction. Using the {\it FORS2} instrument at the 8.2 m {\it VLT}, this area would be covered by $3.2\times 10^4$ pixels. The total error budget here consists of flatfielding errors on the order of 3\% of the background ($\sigma_{FF}$), low-frequency spatial variations of the background across one frame ($\sigma_{sky}$), temporal variations in the background from frame to frame due to OH emissions from the upper atmosphere ($\sigma_{OH}$) and shot noise due to the random statistical fluctuations of the photon flux arriving at each pixel ($\sigma_{SDOM}$). The latter error term is simply the standard deviation of the mean (SDOM) of the flux over an integration bin. $\sigma_{FF}$ is a composite error, which includes residual background unevenness after flatfielding, shutter shading effects and the effects of sensitivity changes as a function of colour across the CCD.

The OH emissions from the ionosphere do not contribute significantly to the $V$ band flux, therefore the total error in this filter is assumed to be \begin{math}
\sigma_V=\sqrt{\sigma_{FF}^2+\sigma_{sky}^2+\sigma_{SDOM}^2}
\end{math}
where we set an upper constraint on the total error by treating $\sigma_{sky}$~and~$\sigma_{SDOM}$ as independent of the flatfielding error. 
As we approach the near-IR range, temporal variations of the OH emission will cause a non-negligible increase of the total error budget. This effect is already visible in the optical $I$ band and we account for it by including an extra term in the total error for that filter:
\begin{math}
\sigma_I=\sqrt{\sigma_{FF}^2+\sigma_{sky}^2+\sigma_{OH}^2+\sigma_{SDOM}^2}
\end{math}
where $\sigma_{OH}$ reflects the change in the background flux due to the variation of the OH emissions from frame to frame. To estimate the size of this term we obtained a temporal sequence of surface brightness variations of the background in the $I$ band by observing a relatively empty sky field for the duration of 90 minutes on a clear photometric night with the 2.52 m {\it Nordic Optical Telescope} and the {\it ALFOSC} instrument. Although the baseline is not particularly long, a distinct periodic shape of the variations was detected. A typical variation of the OH emission in one period was measured to be 2\% of the background -- a factor 24 greater than the shot noise in the same data set. The $\sigma_{OH}$ term can in principle be kept small if one has access to an instrument with very fast readout times, but for our immediate purpose we assume the term cannot be neglected. Allowing a maximum total error per filter of $0.1$ mag we arrive at a required $S/N$ per pixel of 0.076 in the $V$ band and 0.175 in the $I$ band. To reach the surface brightness levels of $\mu_V\approx28.8$ mag arcsec${}^{-2}$ and $\mu_I\sim27.1$ mag arcsec${}^{-2}$ discussed in Sect.~\ref{discussion} we would need 2.2 h for the $V$ band and 4.0 h for the $I$ band with {\it VLT/FORS2}.

\subsection{Near-IR surface photometry}
If the red excess is assumed to appear at isophotes as bright as $\mu_J
\sim 25.2$ mag arcsec$^{-2}$, the required observations can be performed
from the ground assuming they are photon-limited. Groundbased
observations in the NIR regime are completely background-dominated, so
the readout noise can be safely ignored. To comply with the requirement that the total error in each filter should not exceed $0.1$ mag over each integration area of  $3.3$ arcmin times $10$ arcsec, we thus need a signal-to-noise ($S/N$) per pixel of $\sim4\times 10^{-4}$. With the 8.2 m {\it VLT} and the {\it HAWK-I} instrument, we can then reach $\mu_J\sim 25.2$ mag arcsec$^{-2}$ in a total of $\sim3$ hours. To push the surface brightness to even fainter levels we would, however, need to resort to space-based observations. The 6.5 m {\it JWST}, tentatively scheduled for launch in 2018, appears to be the most efficient choice. To estimate the exposure times for the {\it NIRCam} instrument (field of view $2.2\arcmin \times 2.2\arcmin$), we estimate the increase in $S/N$ due to the integration area ($2.2\arcmin$ times 10\arcsec) and compare to the available $S/N$ measurements per resolution element for point sources published on the NIRCam website. We assumed a total system transmission of $\sim0.4$, a readout noise of $15$ electrons per readout, and a maximum of $20$ readouts (destructive, array is reset) for the chopping mode. The readout noise per readout clearly dominates the total background noise over the integration bin as estimated from the available wavelength-dependent scattered light measurements at the NIRCam website, so we have neglected the background noise contribution to the total error budget (the readout noise for one readout is a factor of $\sim50$ greater than the background scattered light noise in the $J$ band). The total error we obtain is therefore much smaller than the required $0.1$ mag per filter, so we take $S/N\sim10$ to be conservative and arrive at a maximum total exposure time of $2$ hours in order to reach surface brightness levels of 27.7 mag arcsec$^{-2}$ (Sect.~\ref{surfphot}) in the $J$ band.

\end{document}